\def\etal{{\rm et al. }}
\def\mpc{{h^{-1} ~ \rm Mpc}}

\def\kms{{\rm km ~ s^{-1}}}
\newcommand\aap{{\em A\&A}}

\newcommand\aj{{\em AJ}}
\newcommand\apj{{\em ApJ}}
\newcommand\apjl{{\em ApJ}}
\newcommand\apjs{{\em ApJS}}

\newcommand\mnras{{\em MNRAS}}
\newcommand\nat{{\em Nature}}

\documentclass[usenatbib]{mn2e}

\usepackage[dvips]{graphicx}

\begin{document}


\title[Triplets of Quasars at high redshift I: Photometric data]{Triplets of 
Quasars at high redshift I: Photometric data 
}
\author[Alonso et al.]
{
\parbox[t]{\textwidth}{
M. Victoria Alonso$^{1}$, Georgina V. Coldwell$^{1}$,
Ilona S{\"o}chting$^{2}$, Carlos Bornancini$^{1,3}$,
Malcolm G. Smith$^{4}$, Diego Garc\'ia Lambas$^{1}$ \& Armin Rest$^{4,5}$}
\vspace*{6pt} \\ 
$^1$ IATE CONICET-Observatorio Astron\'omico de C\'ordoba, C\'ordoba, 
Argentina.\\
$^2$ University of Oxofrd, Astrophysics, Denys Wilkinson Building, 
Keble Road, Oxford OX1 3RH, UK\\
$^3$ Fellow of Secretar\'ia de Ciencia y T\'ecnica de la Universidad 
Nacional de 
C\'ordoba, C\'ordoba, Argentina.\\
$^4$ Cerro Tololo Inter-American Observatory (CTIO), Colina
el Pino S/N, La Serena, Chile\\
$^5$ Physics Department, Harvard University, 17 Oxford
Street, Cambridge, MA 02138}

\date{Accepted 2008 January 21. Received 2007 December 3}

\pagerange{\pageref{firstpage}--\pageref{lastpage}}

\maketitle

\label{firstpage}

\begin{abstract}
We have conducted an optical and infrared imaging in the neighbourhoods
of 4 triplets of quasars. R, z$^\prime$, J and K$_s$ images were obtained with
MOSAIC II and ISPI at Cerro Tololo Interamerican Observatory. Accurate
relative photometry and astrometry were obtained from these images for 
subsequent use in deriving photometric redshifts.
We analyzed the homogeneity
and depth of the photometric catalog by comparing with results coming 
from the literature.
The good agreement shows that our magnitudes are reliable to study
large scale structure 
reaching limiting magnitudes of  R = 24.5, z$^\prime$ = 22.5, 
J = 20.5 and 
K$_s$ = 19.0.  With this catalog we can study the neighbourhoods of the
triplets of quasars searching for galaxy overdensities such as groups and 
galaxy clusters.

\end{abstract}


\section{Introduction}

Our present understanding of galaxy formation and evolution can
be improved significantly through studies of quasars.
In the unified scenario, merger-driven processes include the origin of a
quasar by self-regulating, supermassive, black-hole growth \citep{hop}.  
Thus, most massive spheroids should host an optically-bright
quasar during a brief period of its evolutionary history.

There is well-established evidence indicating that several galaxy
properties  depend on the environment where the galaxies formed and evolved.
Several processes - such as stimulated or truncated star formation,
tidal stripping, merging, etc. - 
can determine the nature of galaxy populations and dynamics.
Also to be considered is the possibility that AGN feedback processes
could induce significant changes in the evolution of
galaxies near to the quasar (see for instance \citet{croton}).
By studying quasar environments we may deepen our understanding of the
relation of quasar fueling and galaxy interactions and merging; and as
a consequence, the formation of bright galaxies.

The availability of large quasar samples has increased significantly in 
recent years, mainly 
due to the advent of large surveys such as the
2dF Galaxy Redshift Survey \citep{colles} and the Sloan Digital Sky Survey
\citep{stou}.  These samples provide an opportunity for more robust 
statistical studies of the quasar phenomenon and its link with galaxy 
formation.

At low redshifts (z $\le $ 0.3), existing 
cross-correlation analysis of
quasar environments show that quasars inhabit
environments similar to those of normal galaxies (\citet{smitha}, 
\citet{cold02}. However, on small scales
(projected distance, $r_p \leq 1 \mpc$; and radial velocity difference, 
$\Delta V = 500 \kms$), the 
quasar environment is
overpopulated by blue disk galaxies having a strong star-formation rate with
respect to typical galaxy neighbourhoods
\citep{cold03,cold06}. This result is in 
agreement
with those of \citet{ilona02,ilona04} in the sense that
 low-redshift
quasars follow the Large Scale Structure traced by galaxy clusters but
they are not located near the center of clusters.  These quasars are
  mainly found in the periphery of such structures
or between two, possibly merging, clusters.

At higher redshifts, quasars are often associated with rich environments
\citep{hall,djor} where the comoving density of
galaxies  is higher than
that expected for the general field.  This should be interpreted as the
core of future rich clusters.  Some results at $z \sim 1$
suggest that quasars reside on the peripheries of clusters or in cluster
mergers \citep{haines01,haines04,tanaka00,tanaka01}.
\citet{tanaka01}, for example, investigated a group of 5 quasars
tracing a $\sim$ 10
Mpc structure of 4--5 clusters, with only a single
radio-loud quasar appearing to be directly associated with one
of the clusters.

In general, the existing studies of quasar environments suggest their 
association with
forming structures, marking merging clusters and filaments. Consequently,
groupings of quasars can be expected to trace regions of extraordinary
activity.
Pairs of quasars have been strongly studied and they are used to identify
cluster of galaxies at high redshifts. \citet{zhdanov} 
suggested that
quasar pairs can be used as tracers of high redshift large-scale structures.
Moreover, \citet{djor03} found a quasar pair at  $z \sim 5$
associated with 
large scale structure, possibly a protocluster at that redshift. More 
recently, \citet{boris}
found that quasar pairs at $z \sim 1$ are excellent tracers of high 
density environments suggesting 
that they can be used to find galaxy clusters.

Triplets of quasars in a relative small volume are extremely rare and 
until now, no cluster-scale triplets have been published or investigated.
A very compact quasar association ($\sim$ 50 Kpc separation) 
was recently reported by \citet{djor07} suggesting that this could be a compact galaxy group 
in a merging process.
We propose to use triplets of 
quasars as ``lighthouses'' to find and study clusters of galaxies that favour
 such rare events. With this aim in mind, we 
performed an optical and infrared photometric study of the environments of
triplets of quasars at different redshifts 
on the scale of galaxy clusters. The images allow us 
to observe galaxies in the early Universe and
they are used to investigate the environments of quasar triplets
We also
study the group and cluster morphology, and the photometric properties 
of individual galaxies in the 
neighbourhood of each triplet. We have already found that low-z triplets
of Seyfert 1 galaxies are 
associated with extremely rich 
clusters of galaxies \citep{ilona07}.  If the quasar triggering 
mechanism does not evolve strongly over redshift we can expect to find some of 
the richest galaxy clusters ever discovered at $z \sim 1$. A negative
result will be clear evidence of the evolution of the quasar formation 
mechanism.

This paper is organized as follows:
In section 2 we describe the sample selection and section 3 shows the
optical and infrared observations and data
reduction. Section 4 describes the photometric galaxy catalog,
and in section 5 we provide a summary of the main results.
We adopt the latest cosmology with $\Omega_M = 0.3$, $\Omega_{\lambda} = 0.7$
and a Hubble constant, $H_0=70 ~ {\rm{km ~ s^{-1} ~ Mpc^{-1}}}$.

\section{The Sample}

The quasar sample is defined by quasars which have at least one emission 
line with a full
width at half maximum larger than 1000 $\kms$, luminosities brighter than 
$M_i = -23$,
PSF $i$ magnitude $ < 19.1$ and highly reliable redshifts \citep{schneider03}.
To analyze the regions in the neighbourhoods of groups of quasars, we 
identified different active systems
using the friend-of-friend, three-dimensional, percolation algorithm 
(\citet{huchra}). This technique was applied to the sample
of 54000 quasars taken from SDSS-DR4 which satisfy the criteria defined
above.  Taking into account galaxies in clusters and their surroundings, 
and the
low spatial density of quasars (about 15 quasars per square degree), 
we finally defined as candidates for triplets of quasars 
those systems with percolation
 longitudes smaller than
2 $\mpc$ and 2000 $\kms$.  These limits are about twice the typical region 
for rich galaxy clusters.  We found 293 pairs of
quasars, only 7 triplets and no systems with more members. To increase
the sample of triplets, we applied the same criteria to the
\citet{veron} catalogue. 

\begin{figure}
\includegraphics[width=72mm,height=75mm]{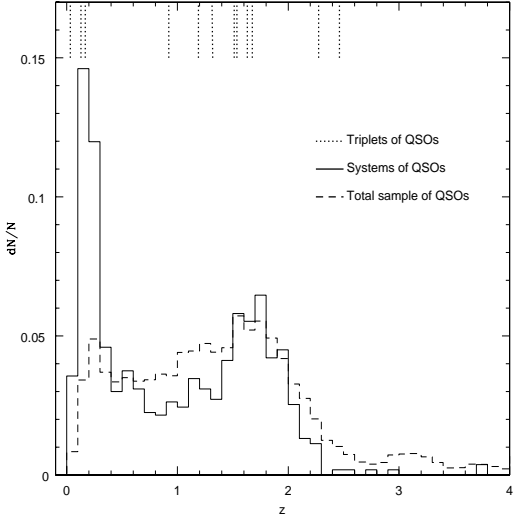}
\caption{Relative
redshift distribution of the total quasar sample (dashed line), and systems of
quasars represented by pairs and triplets (solid line).  The redshifts of the 12 triplets in 
our sample 
are marked at the top of the figure.}
\label{fig1}
\end{figure}

Figure~\ref{fig1} shows the 
redshift distribution of the fraction of the quasar number normalized to 
the total quasar sample, and that corresponding 
to pairs and triplets (systems) of quasars 
found with the adopted selection criteria.  The redshifts of each of the twelve triplets in the total sample 
of triplets are marked at the top 
of the figure. For systems of quasars, there are two 
peaks at z $\sim$ 0.2 and z $\sim$ 1.8, which are
more concentrated than the total
quasar sample.  In spite of the uncertainties due to the small amount of data 
for systems of quasars, the presence of 
these two peaks may be interpreted
in terms of differences in the growth of self-regulated black holes for
different masses \citep{dimatteo}.
They proposed a fast evolutionary growth for the massive black holes formed in the
early Universe, which exhausted the surrounding gas due to a higher accretion rate.
On the other hand, the lifetime of the active phase increases for
black holes of smaller masses implying an excess of low luminosity quasars
in the local Universe.  

The total sample
consists of 12 triplets of quasars: 3 triplets with z $<$ 0.2 and 9
within the range of 0.9 $<$ z $<$ 2.5.  As mentioned above, pairs of 
quasars - as a class - have been extensively studied and so we will
concentrate our efforts here only on triplets of quasars, which are very rare.

Our first subsample of triplets of quasars, those nearby triplets with 
z $<$ 0.2 is discussed
in \citet{ilona07}.  Using the photometric and spectroscopic 
catalogs of
the SDSS-DR4, we found a strong association of z $<$ 0.2 triplets with the 
richest central parts of superclusters. In all cases the individual members of the 
triplets have been found on the periphery of the main galaxy concentrations.

For some of our second subsample, those triplets at higher redshifts, 
we performed
multicolor photometry of the regions near to these triplets in order to locate
associated galaxy groups and clusters.

\section{Observations and data reduction}

In this paper, we concentrate on four high-redshift triplets from the 
sample. 
We obtained multicolor photometry of their neighbourhoods 
with the CTIO 4m telescope using two instruments: MOSAIC II for the optical and ISPI for
the near IR (Proposal Ids. 06A-0146 and 
06B-0117). Table 1 shows the observed
triplets, including the triplet and quasar identifications, coordinates, 
redshifts and absolute V magnitudes, respectively.

\subsection{Optical imaging}

Triplets 5 and 6 were observed in the R
and z$^\prime$ bands using MOSAIC II during 5 half nights from February 2$^{nd}$ through 
6$^{th}$,
2006.  This optical instrument is an array of 8 2048 $\times$ 4096 SITe CCDs 
with a gain of 2 e-/ADU and a scale of 0.27 arcsec/pixel, giving a total 
field of view of 36 $\times$ 36 arcmin$^2$.  In the R band, we made eight
 exposures of 600s for Triplet 5 and eleven exposures of 600s for Triplet 6.
Moreover, four exposures of 600s were taken in z$^\prime$ band for both 
triplets. Table 2 shows the triplet Ids together with a summary of the 
optical observations, in R and z', in columns 2 and 3, respectively.

\subsubsection{Photometry}

The images were reduced using the
SMSN pipeline \citep{Rest05,Garg07,Miknaitis07} originally created for
the SuperMACHO and ESSENCE projects. The SMSN pipeline reduces the raw
images using the IRAF \footnote{Image Reduction 
and Analysis Facility (IRAF), a software system 
distributed by the National Optical
Astronomy Observatories (NOAO)} mosaic routines XTALK and MSCCMATCH, as well as
its own routines in C, perl and python. As part of this reduction, the
images are crosstalk
corrected, astrometrically calibrated, and split into the 16 separate
amplifiers.  The images are then bias subtracted and flat fielded. The
z-band images were also defringed.

We flux calibrated the R and z$^\prime$
images using our observations of several  \citet{landolt} and \citet{smithb} 
standard stars, respectively. 
In our images, point-source magnitudes were obtained using SExtractor 
\citep{bertin} 
with an aperture diameter of $6\arcsec$, 
which proved to be adequate by monitoring the growth curve of all the 
measured stars.  For the final calibration, we include the zero point and
extinction term.  The colour term was negligible compared to magnitude errors 
and it was not included (as in \citet{gawiser}).

\begin{figure}
\includegraphics[width=72mm,height=75mm]{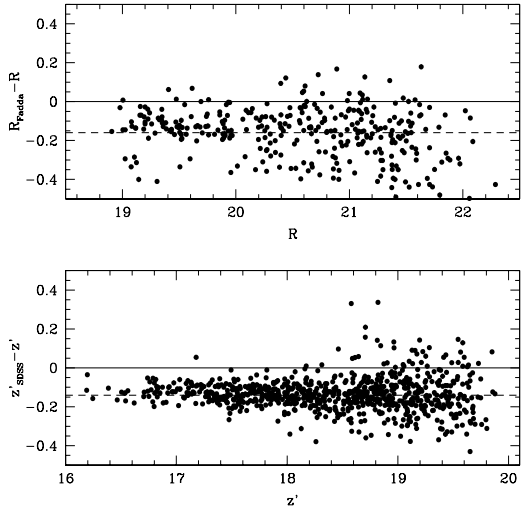}
\caption{Magnitude comparisons for optical bands between stars in our 
observed fields and
those stars in common with SDSS.  Upper panel shows the 
R-band magnitude differences before applying Fadda et al (2004) 
transformation to SDSS r$\arcmin$ magnitudes.  Bottom panel shows the z$^\prime$-band 
magnitude differences.}
\label{fig3a}
\end{figure}

For R magnitudes, the zero point is in agreement with the published value
for the instrument.  In order to better analyze the precision of our
calibration, 
we also compared the magnitudes of the stars in our fields obtained with 
SExtractor ({\tt MAG\_BEST}) with the PSF magnitudes computed by SDSS.
 Before performing the comparison we transformed SDSS r$\arcmin$
magnitudes to our system using Fadda et al. (2004) transformations.
Figure~\ref{fig3a} shows in the upper panel this comparison of R magnitudes 
showing the good agreement with those of the SDSS.  The mean difference is 
$\Delta$ R = 0.162 
$\pm$ 0.139.

In the z$^\prime$-band the comparison with SDSS was made without any 
transformation because
the filters are identical.  The bottom panel of Figure~\ref{fig3a} shows this 
comparison.  The
mean difference is $\Delta$ z$^\prime$ = 0.14 $\pm$ 0.08 mag. This offset 
 is consistent with the estimated uncertainty of the overall
zero-point, mainly due to the very small number of standard stars used in this 
band.  Our final z$^\prime$ magnitudes were corrected by this 
offset.

We applied the offsets in both bands to correct our final magnitudes. Note also
that all magnitudes are in the Vega system.

\subsubsection{Astrometric precision}

\begin{figure}
\includegraphics[width=72mm,height=75mm]{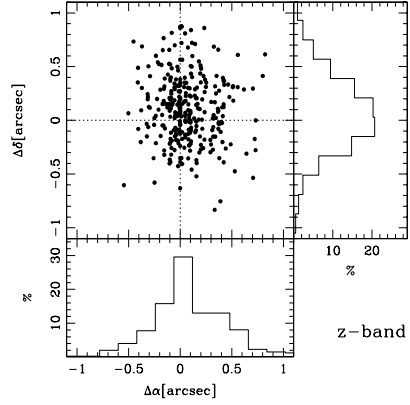}
\caption{Astrometric differences in coordinates between the matched MOSAIC II
 and USNOA2.0
stars in the z$^\prime$-band.}
\label{fig2}
\end{figure}

In order to assess the accuracy of the astrometric calibration, we compared
the position of the stars in the observed fields with those of the
USNO-A2.0 Catalog (Monet \etal 1998).  We used only unsaturated stars 
from our images. Figure~\ref{fig2} shows only the results of the comparison
for the z$^\prime$-band.
The mean offset in the positions between both catalogs is 
negligible: $\Delta \alpha = -0.057 \pm 0.281$ arcsec and 
 $\Delta \delta = -0.097 \pm 0.302$ arcsec for R-band and 
$\Delta \alpha = -0.110 \pm 0.236$ arcsec and 
 $\Delta \delta = -0.074 \pm 0.302$ arcsec for z$^\prime$-band.
The scatter in the
coordinates is found to be typically  smaller than 0.3 arcsec and it 
is consistent
with the astrometric accuracy of the USNO-A2.0 catalog.
We also compared the positions of the stars with SDSS sources 
and we found systematic offsets of 0.218 $\pm$ 0.096 and 0.434 $\pm$ 0.107 
in right ascension and declination, respectively - similar to those found by 
Fadda \etal (2004).

\subsection{Infrared imaging}

The triplets of quasars were also observed in the infrared using ISPI in two 
runs during 2006:  
5 half nights from 
April 2$^{nd}$ to 6$^{th}$ and 3 complete nights - October 7$^{th}$ to 
9$^{th}$.  
ISPI (Infrared Side Port Imager) is a 1-2.4 $\mu$m imager
with an 2048 $\times$ 2048 HgCdTe array with an scale of 0.3 arcsec/pix,
giving a field of view of 10.5 x 10.5 
arcmin$^2$ \citep{ispi}.  The gain is 4.25 e-/ADU and the 
dark current is very low - about 0.1 e-/sec for long exposures.
However, this array has a number of bad pixels which are located along a 
few rows and columns in the upper right corner.

All four triplets were observed in the K$_s$ band but only Triplets 6 and 7 were observed 
in the J band.  We used a circular dithering pattern of 15 different 
positions to observe the triplet fields.  For each image in the sequence, 
the exposure 
time was 130s $\times$ 1 coadds and 20s $\times$ 5 coadds for J and K$_s$,
respectively.  Table 2 includes a summary of the 
IR observations in columns 2 to 5.

\subsubsection{Photometry}

For the reduction, we used standard tools in IRAF 
 and some tasks from the CTIO Infrared reduction package ({\tt cirred}) 
written by R. Blum for specific ISPI processing. The basic data reduction 
consisted in creating the master flatfield image from a median of  
lamps on -- lamps off dome flats and the construction of the bad pixel mask.  
The final 
master flat was corrected by illumination effects and normalized by the 
average intensity.  Median sky images were created for each dithered sequence.
The scientific images were flatfielded, bad pixel corrected, and sky 
subtracted. The astrometry was performed with the package WCSTools 
\footnote{Available at  ftp://cfa-ftp.harvard.edu/pub/gsc/WCSTools/} 
\citep{wcs}.  Then
for each sequence the images were weighted combined with individual weight 
map images using {\tt SWarp} written by E. Bertin \footnote{
http://terapix.iap.fr/soft/swarp}.

To calibrate the magnitudes, we observed some standard stars from Persson 
et al. (1998).  We also used those stars in our fields belonging to the 2MASS 
point source catalog 
\citep{cutri} to improve precision. As for the optical magnitudes, the
stellar instrumental magnitudes were obtained with an aperture diameter setting of 
$6\arcsec$.  For the final calibration, again we did not include the
color term.

\begin{figure}
\includegraphics[width=72mm,height=75mm]{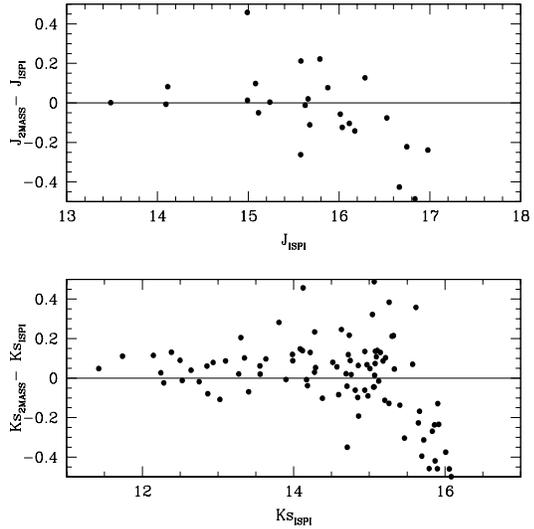}
\caption{Magnitude comparisons for IR bands between stars in
common with 2MASS. Upper and bottom panels show magnitude differences vs 
our magnitudes for J and Ks bands, respectively.}
\label{fig5}
\end{figure}

To verify the photometric precision we compared our
magnitudes 
of detected objects (those with 5 $\sigma$ above the sky background) 
with the corresponding, uncontaminated, point sources from the 2MASS catalogue.  
Figure~\ref{fig5} presents  comparisons of the magnitude 
differences 
for the stars in common with 2MASS in J and K$_s$ bands, respectively; the agreement is excellent. 
In the J band, for objects brighter than 16 mag, 
$\Delta J = -0.007 \pm 0.153$ mag.
In the K$_s$ band, the comparisons look similar; 
$\Delta K_s = -0.066 \pm 0.109$ mag for objects brighter than 14.5 mag.
Finally, in both bands, we can see from Figure~\ref{fig5} a
progressive increase in the scatter at fainter 
magnitudes due to larger errors in the 2MASS catalogue at these magnitudes.

\subsubsection{Astrometric precision}

In order to verify the precision of our astrometry, we also checked our measured positions
for 2MASS objects in our fields against the 2MASS catalogue.  Figure~\ref{fig7} shows the 
differences in both coordinates, right ascension and 
declination, 
between the two sets of data in
K$_s$ band. The average differences are 
$\Delta \alpha = 0.096 \pm 0.396$ 
arcsec and $\Delta \delta = 0.006 \pm 0.347$ arcsec for J band for objects 
brighter than 16 mag; and
$\Delta \alpha = 0.038 \pm 0.313$ arcsec and 
$\Delta \delta = 0.018 \pm 0.371$ arcsec for K$_s$ band for objects brighter 
than 14.5 mag, an overall astrometric precision sufficient for our 
purposes.  

\section{Source detection and catalogue}

Our photometric catalog in the four bands was constructed using SExtractor.
An object was considered detected if the flux had an
excess of 1.5 times the local background noise level over at least 
5 connected pixels.

The catalog includes magnitudes in apertures 
of 2, 3 and 4 
arcsec, total magnitudes, star-galaxy separation, ellipticity, etc.  The
star-galaxy classification
 represented by the stellarity index, {\tt CLASS\_STAR} 
allows us to decontaminate the galaxy catalog. 
Figure~\ref{fig6} shows this
parameter as a function of R magnitudes. The results are similar for 
z$^\prime$ band.  From the Figure we can see clearly
two object sequences: stars defined with {\tt CLASS\_STAR} $\sim$ 1
and galaxies with {\tt CLASS\_STAR} $\sim$ 0.  At fainter magnitudes, 
Bertin \& Arnouts (1996) showed that the stellarity index is strongly 
seeing dependent.  Taking this into account, we defined as
galaxies those objects with {\tt CLASS\_STAR} $<$ 0.9 and 
{\tt CLASS\_STAR} $<$ 0.75  for R and z$^\prime$ bands, respectively.

The IR photometric catalog was constructed using SExtractor in double-image
mode with the K$_s$ image as a reference on account of its better quality. 

Figure~\ref{fig7} shows the IR colour-magnitude diagram and how 
the $J-K_s$ colour alone provides
a reasonable means of performing the star-galaxy separation.
The horizontal sequence at $J-K_s < $ 1 corresponds to the stellar 
sequence of galactic stars
with types later than G5 and earlier than K5 \citep{finlator}.
In the Figure 
we also plot point sources detected with 
{\tt CLASS\_STAR} $>$ 0.95  (star symbols). As can be seen most of these
objects are located within the stellar locus box. But those point sources with 
redder colours ($J - K_s > $1 mag and $K_s > $ 18) are too faint and the 
star-galaxy separation 
becomes unreliable. In spite of this, we prefer to include them in the final 
catalog (also using star symbols).  

To reinforce our star-galaxy separation we also use 
the half-light radius parameter, $r_{1/2}$. This parameter 
 measures the radius that encloses 50\% of the object's total flux. 
It is clearly independent of magnitude and 
depends only of the image seeing.
In Figure~\ref{fig7a} we show the half-light radius vs $K_s$ magnitude. Again,
as in Figure~\ref{fig7} point sources 
with {\tt CLASS\_STAR} $>$ 0.95 are plotted with star symbols. 
As mentioned in Sub-Section Infrared imaging, there are some 
bad pixels located along a few rows and columns and yet they were not 
rejected during the reduction process. These objects have 
 $r_{1/2} < $ 1 in the Figure and they were discarded in the final catalog.
Also, the stellar 
locus is clearly visible at $r_{1/2} \sim $ 2 and $K_s < $ 18. Therefore, 
we define as galaxies those objects with $J-K_s > $ 1 and $K_s < $ 18 mag.

We created the final photometric catalog discarding spurious objects and
stars.

\subsection{Results. Number counts}

The final catalog in all bands was obtained with total magnitudes, and no
need to apply an aperture correction.  Our fields are located at 
$\mid b \mid > 40  $\degr, so the Galactic extinction effect is generally small.  Only
the R magnitudes were corrected for galactic extinction 
using the $E(B-V)$ values provided in the NASA / IPAC Extragalactic Database 
(NED) \footnote{http://nedwww.ipac.caltech.edu/ - the NASA-IPAC Extragalactic 
Database}, which are based on the $E(B-V)$ values from the extinction maps of 
\citet{sch}.  

The number counts as a function of magnitude allow us to analyze the 
homogeneity
and depth of the catalog and to compare it with results obtained from
the literature.

Table 3 shows the main result, the number of galaxies per square degree per magnitude
in the four bands: R, z$^\prime$, J and K$_s$ and their respective errors.
The Figures~\ref{fig8}, \ref{fig8a}, \ref{fig8c} and 
\ref{fig8b} show the galaxy 
number counts, in the four bands.  We compared our results with galaxy number counts 
from the literature, mainly: 
MacDonald et al. (2004) and Metcalfe et al. (2001) in the R band; 
Kashikawa et al. (2004) and Capak et al. (2004) in the z$^\prime$ band; 
Olsen et al. (2006), Saracco et al. (1999) and Iovino et al. (2005) for
the J band; and finally 
Best et al. (2003), Olsen et al. (2006), Best (2000) and Bornancini
et al. (2004) for K$_s$ band.  From these Figures it is clear that 
we reach magnitudes of 
R = 24.5 mag, z$^\prime$ = 22.5 mag, J = 20.5 mag and 
K$_s$ = 19.0 mag.  Within these limiting magnitudes, we 
find good agreement with the galaxy counts
coming from other authors in all bands.


\section{Summary}

We selected a sample of triplets of quasars as those triple systems with 
percolation
 longitudes smaller than
2 $\mpc$ and 2000 $\kms$.  The goal is to use them as ``lighthouses'' to find 
richest clusters. For this purpose we conducted a photometric study of the 
neighbourhoods of
 4 triplets of quasars in the redshift range of 0.9 to 1.7. Images in R, 
z$^\prime$, J and K$_s$ bands were obtained with
MOSAIC II and ISPI at CTIO.  Magnitudes and coordinates were found to be in excellent agreement with
the USNO-A2.0 and 2MASS catalogs. 
We analyzed the homogeneity
and depth of the photometric catalog by comparing the number counts 
with results coming from the literature.  
The excellent agreement shows that our magnitudes are reliable, 
down to limiting magnitudes of R = 24.5, z$^\prime$ = 22.5, 
J = 20.5 and K$_s$ = 19.0.  These results allow us to study large- 
scale 
structure and in a forthcoming paper we will
deal with the statistical
analysis of this data.

\section{Acknowledgments}

This research was partially supported by grants from  CONICET,
Agencia C\'ordoba Ciencia 
 and the  Secretar\'{\i}a de Ciencia y T\'ecnica
de la Universidad Nacional de C\'ordoba.  GVC would like to
thank the CTIO for hospitality during her observational runs.
IKS would like to acknowledge the kind hospitality of the IATE group during her
stay at Cordoba. We would also like to thank Roger Clowes for sharing with 
us his fringe frames for MOSAIC data.
This publication makes use of data products from the Two Micron All Sky 
Survey, which is a joint project of the University of Massachusetts and the 
Infrared Processing and Analysis Center/California Institute of Technology, 
funded by the National Aeronautics and Space Administration and the National 
Science Foundation. Also, the authors make use of the NASA/IPAC extragalactic 
database (NED) which is 
operated by the Jet Propulsion Laboratory, Caltech, under contract with the 
National Aeronautics and Space Administration.

{}

\begin{table}
 \caption{The Observed Sample of Triplets of Quasars }
 \label{tab:trips}
{\small
 \begin{tabular}{lrcll}
\hline
Triplet      &    RA(J2000) &   DEC(J2000) &  z  &   M$_V$ \\
Quasar Id.      &              &              &     &           \\
\hline
\hline
             &              &              &     &           \\ 
TRIPLET 4    &              &              &     &           \\  
          2QZ J015646-2828  &  1:56:46.6  & -28:28:0   &  0.919  & -24.5 \\
          2QZ J015647-2831  &  1:56:47.9  & -28:31:43  &  0.919  & -24.4 \\
          2QZ J015656-2823  &  1:56:56.0  & -28:23:35  &  0.925  & -24.3 \\
                            &             &            &         &       \\ 
TRIPLET 5                   &             &            &         &       \\  
          SDSS J10023+0155  & 10:02:19.5  &  01:55:37  & 1.509   & -24.8 \\
          SDSS J10025+0150  & 10:02:34.3  &  01:50:10  & 1.506   & -25.7 \\
          SDSS J10026+0159  & 10:02:36.7  &  01:59:48  & 1.516   & -24.7 \\
                            &             &            &         &       \\  
TRIPLET 6                   &             &            &         &       \\  
         2QZ J120914+0035   & 12:09:14.9  &  00:35:51  & 1.316   & -25.8 \\
         2QZ J120919+0029   & 12:09:19.6  &  00:29:26  &  1.319  & -24.3 \\
         2QZ J120922+0026   & 12:09:22.4  &  00:26:46  & 1.322   & -24.2 \\
                            &             &            &         &       \\ 
TRIPLET 7                   &             &            &         &       \\  
          2QZ J235405-2845  &  23:54:05.3 &  -28:45:06 &  1.673  & -25.2 \\
          2QZ J235430-2848  &  23:54:30.2 &  -28:48:41 &  1.670  & -24.6 \\
          2QZ J235414-2848  &  23:54:14.2 &  -28:48:10 &  1.679  & -25.8 \\
\hline
\hline
  \end{tabular}
}
\end{table}

\begin{table}
 \caption{Observing Logs}
 \label{tab:tablog}
{\small
 \begin{tabular}{lrccc}
\hline
Triplet Id.     &  R & z$^\prime$ & J & K$_s$\\
\hline
\hline
                            &     &    &    &       \\
TRIPLET 4                   &  -- & -- & -- & 7 $\times$ 15 $\times$ 5 $\times$ 20s \\  
TRIPLET 5                   &  8 $\times$ 600s & 4 $\times$ 600s & -- & 8 $\times$ 15 $\times$ 5 $\times$ 20s \\  
TRIPLET 6                   & 11 $\times$ 600s & 4 $\times$ 600s & 15 $\times$ 15 $\times$ 130s & 8 $\times$ 15 $\times$ 5 $\times$ 20s \\  
TRIPLET 7                   &  -- & -- & 12 $\times$ 15 $\times$ 130s & 7 $\times$ 15 $\times$ 5 $\times$ 20s \\  
\hline
\hline
  \end{tabular}
}
\end{table}

\begin{table}
\caption{Number Counts in N Deg$^{-1}$ 0.5 Mag$^{-1}$}
 \label{tab:tabcounts}
{\small
 \begin{tabular}{lcccc}
\hline
Magnitude    &  R & z$^\prime$ & J & K$_s$\\
\hline
14.75  & ...            &    ...         &     ...         & 126 $\pm$63    \\
15.25  & ...            &  ...           &   ...           & 347 $\pm$104   \\
15.75  & ...            &  ...           &   ...           & 663 $\pm$144   \\
16.25  & ...            &  ...           &  410 $\pm$113   &  1359 $\pm$207 \\
16.75  & ...            &   ...          &  726 $\pm$151   & 2117 $\pm$258  \\
17.25  & ...            & 151 $\pm$21    & 979 $\pm$175    &  4013 $\pm$356 \\
17.75  & ...            & 281 $\pm$29    & 2180 $\pm$262   &  5466 $\pm$415 \\
18.25  & ...            & 446 $\pm$36    & 2054 $\pm$254   & 10110 $\pm$565 \\
18.75  & ...            & 814 $\pm$49    & 4013 $\pm$356   & 13680 $\pm$657 \\
19.25  & 357 $\pm$32    & 1489 $\pm$66   &  6098 $\pm$438  & 15580 $\pm$701 \\
19.75  & 804 $\pm$48    & 2269 $\pm$81   &  7520 $\pm$487  & 15580 $\pm$701 \\
20.25  & 1416 $\pm$64   & 3827 $\pm$105  &  11630 $\pm$606 & 13020 $\pm$641 \\
20.75  & 2112 $\pm$78   & 5946 $\pm$131  &  12950 $\pm$639 & 5814 $\pm$428  \\
21.25  & 2945 $\pm$92   & 8487 $\pm$157  &  13110 $\pm$643 & ...            \\
21.25  & 2945 $\pm$92   & 8487 $\pm$157  &  9826 $\pm$557  & ...            \\
21.75  & 4599 $\pm$116  & 14490 $\pm$205 &  6667 $\pm$459  & ...            \\
22.25  & 6754 $\pm$140  & 18820 $\pm$233 &  ...            & ...            \\
22.75  & 10340 $\pm$173 & 17250 $\pm$224 & ...             & ...            \\
23.25  & 16160 $\pm$217 & 10180 $\pm$172 & ...             & ...            \\
23.75  & 26070 $\pm$275 & ...            & ...             & ...            \\
24.25  & 37260 $\pm$329 & ...            & ...             & ...            \\
24.75  & 40550 $\pm$343 & ...            & ...             & ...            \\
25.25  & 11050 $\pm$179 & ...            & ...             & ...            \\
\hline
\hline
  \end{tabular}
}
\end{table}


\clearpage

\begin{figure}
\includegraphics[width=72mm,height=75mm]{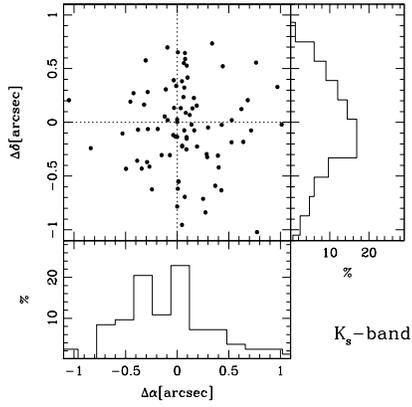}
\caption{Differences in right ascension and 
declination of stars in common with the 2MASS catalog in the
K$_s$ band.}
\label{fig4}
\end{figure}

\begin{figure}
\includegraphics[width=80mm,height=80mm]{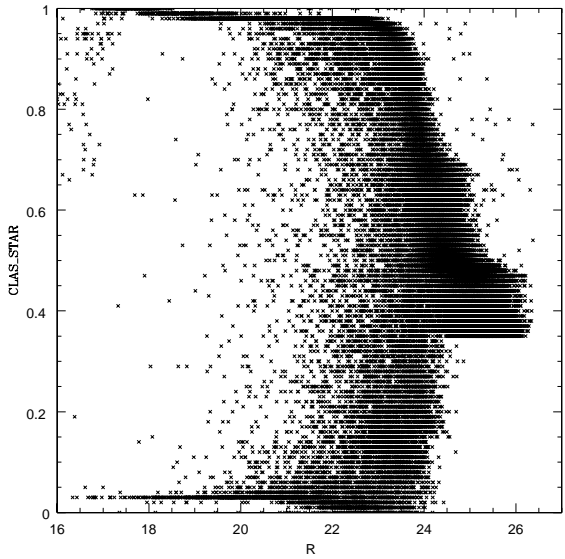}
\caption{The star-galaxy separation versus total magnitudes for R band.}
\label{fig6}
\end{figure}

\begin{figure}
\includegraphics[width=80mm,height=80mm]{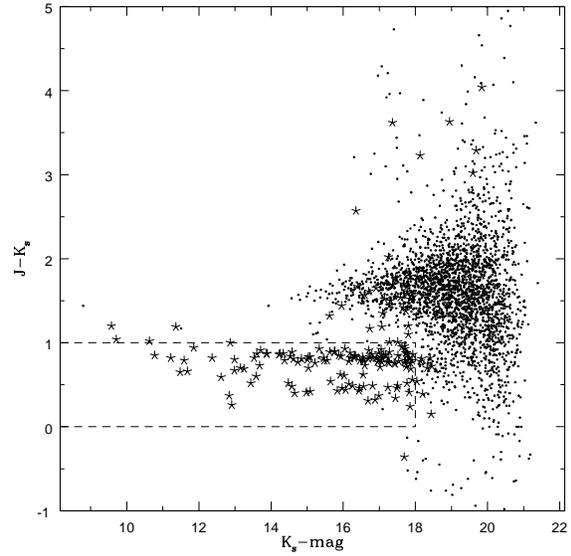}
\caption{IR Color--magnitude diagram to discriminate among stars and galaxies.
Star symbols correspond to those objects with {\tt CLASS\_STAR} $>$ 0.95.}
\label{fig7}
\end{figure}

\begin{figure}
\includegraphics[width=80mm,height=80mm]{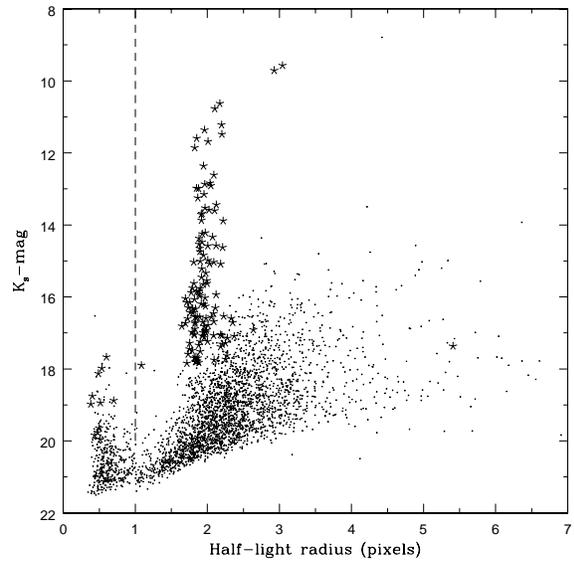}
\caption{Half-light radius vs $K_s$ magnitude. As in previous Figure, star 
symbols correspond to those objects with {\tt CLASS\_STAR} $>$ 0.95.  The 
vertical line marks our detectability limit.}
\label{fig7a}
\end{figure}

\begin{figure}
\includegraphics[width=80mm,height=80mm]{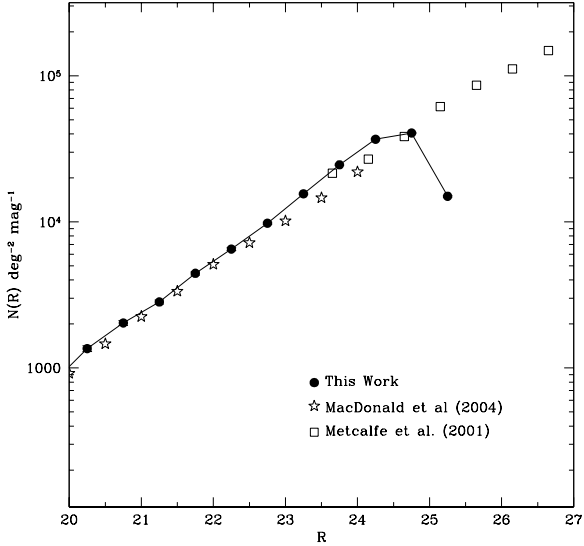}
\caption{Number counts N(R) in the R band.  We also show number counts coming
from other authors: MacDonald et al. (2004) and Metcalfe et al. (2001).}
\label{fig8}
\end{figure}

\begin{figure}
\includegraphics[width=80mm,height=80mm]{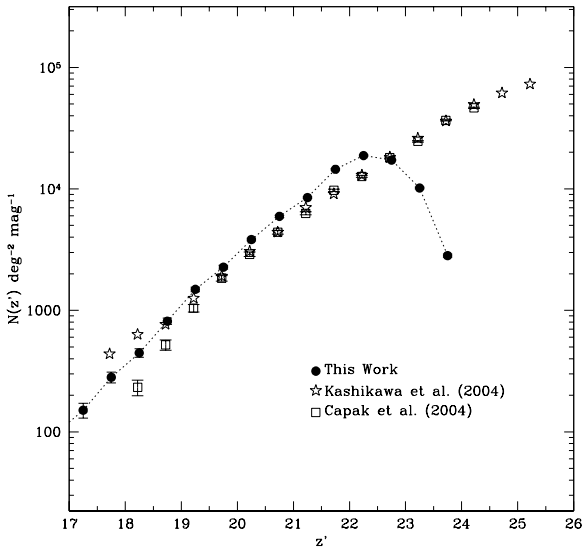}
\caption{Number counts N(z$^\prime$) in the z$^\prime$ band. We also show number counts coming
from other authors: Kashikawa et al. (2004) and Capak et al. (2004).}
\label{fig8a}
\end{figure}

\begin{figure}
\includegraphics[width=80mm,height=80mm]{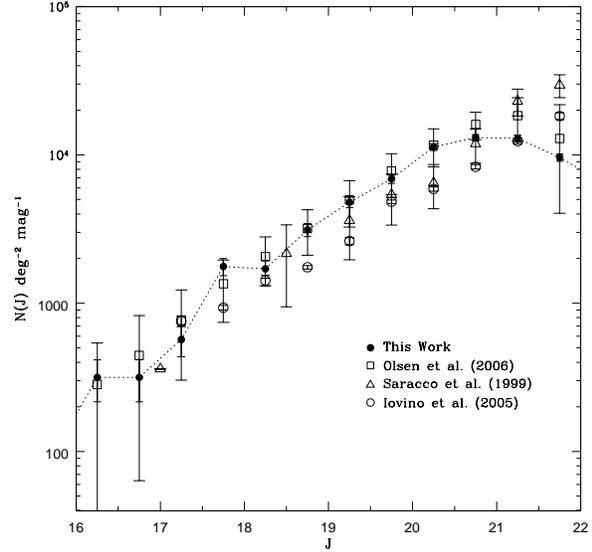}
\caption{Number counts N(J) in the J band. We also show number counts coming
from other authors: Olsen et al. (2006), Saracco et al. (1999) and 
Iovino et al. (2005).}
\label{fig8c}
\end{figure}

\begin{figure}
\includegraphics[width=80mm,height=80mm]{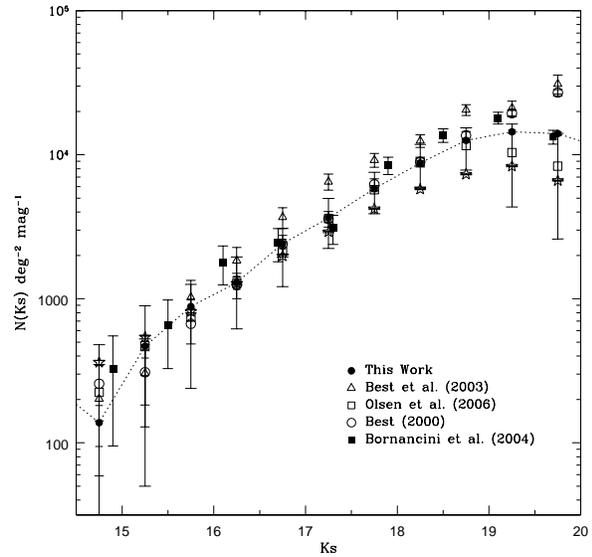}
\caption{Number counts N(K$_s$) in the K$_s$ band. We also show number counts 
coming
from other authors: Best et al. (2003), Olsen et al. (2006), Best (2000) and 
Bornancini et al. (2004).}
\label{fig8b}
\end{figure}

\label{lastpage}

\end{document}